\documentclass[12pt]{article}
 \usepackage{latexsym, amsmath, epsfig, amsfonts, amsbsy}
 \DeclareFontFamily{OT1}{rsfs10}{}
 \DeclareFontShape{OT1}{rsfs10}{m}{n}{ <-> rsfs10 }{}
 \DeclareMathAlphabet{\mathscript}{OT1}{rsfs10}{m}{n}

 \else\oddsidemargin25mm\evensidemargin25mm\marginparwidth25mm\fi
 \def\Z{\mathbb{Z}}

 \def\R{\mathbb{R}}

 \def\bl{\Big\{}
 \def\br{\Big\}}
 \def\bpl{\Big(}
 \def\bpr{\Big)}
 \def\e{\epsilon}
 
 \def\t{\theta}
 
 \def\w{\omega}

 \def\der{\partial}
 \newcommand{\ft}[2]{{\textstyle\frac{#1}{#2}}}
 \def\brr{\begin{equation}}
 \def\err{\end{equation}}
 \def\brr{\begin{eqnarray}}
 \def\err{\end{eqnarray}}
 \def\ba{\left(\begin{array}}
 \def\ea{\end{array}\right)}
 \def\lf{\left.\begin{array}{c}}
 \def\rf{\end{array}\right.}

 \def\cDslash{\hbox{\ooalign{$\displaystyle {\cal D}$\cr$\hspace{.03in}/$}}}

 \def\derbar{\stackrel{\leftrightarrow}{\partial}}
 \newcommand{\dr}{\raise.3ex\hbox{$\stackrel{\leftarrow}{\partial }$}{}}
 \newcommand{\dl}{\raise.3ex\hbox{$\stackrel{\rightarrow}{\partial}$}{}}
 \newcommand{\topi}{\raise.3ex\hbox{$\stackrel{\pi}{\longrightarrow}$}{}}
 \newcommand{\ns}{\normalsize}
 \renewcommand{\theequation}{\arabic{section}.\arabic{equation}}
 \renewcommand{\a}{\alpha}
 \renewcommand{\b}{\beta}
 
 \renewcommand{\d}{\delta}

\begin{document}

 \begin{titlepage}

\title{
   \hfill{\ns HWS-2004A06\\}
   \hfill{\ns DAMTP-2004-76\\}
   \hfill{\ns hep-th/0406152\\[2cm]}
   {\LARGE Central Charges and Extra Dimensions\\
   in Supersymmetric Quantum Mechanics}\\[1cm]}

\author{{\bf
   Michael Faux$^{1}$ ,\,
   David Kagan$^{2}$
   and Donald Spector$^{1}$}\\[5mm]
   {\it $^1$Department of Physics} \\
   {\it Hobart and William Smith Colleges} \\
   {\it Geneva, NY 14456, USA} \\[4mm]
   {\it $^2$DAMTP, Center for Mathematical Sciences}\\
   {\it University of Cambridge}\\
   {\it  Wilberforce Road, Cambridge, CB3 0WA, UK}}

\date{June 2004}

\maketitle

 \begin{abstract}
 \noindent
 We systematically include central charges into supersymmetric
 quantum mechanics formulated on curved Euclidean spaces,
 and explain how the background geometry manifests itself on
 states of the theory.
 In particular, we show in detail how, from the point of view of
 non-relativistic $d=1$ world-line physics,
 one can infer the existence of target space dualities
 typically associated with string theory.  We also explain in
 detail how the presence of a non-trivial supersymmetry
 central charge restricts the background geometry in which
 a particle may propagate.\\[.2in]
\end{abstract}

\thispagestyle{empty}

\end{titlepage}

 \setcounter{equation}{0}
 \section{Introduction}
 As is well appreciated, supersymmetry \cite{susy} is a concept which not only
 provides elegant and useful solutions to interesting problems,
 such as the hierarchy problem in the standard model, but which also
 plays a key role in the structure
 of a variety of theories. For example, it appears as a required
 ingredient in consistent string theories \cite{string},
 and also underlies the presence of shape invariance in exactly
 solvable systems in ordinary quantum mechanics \cite{shapes}.  As
 is also well appreciated, attempts to find more fundamental
 descriptions of nature frequently benefit from the inclusion of
 extra, less obvious, dimensions as part of our physical space.
 It is interesting to consider what the two ideas of
 supersymmetry and extra dimensions imply, at a basic level, when
 they are imposed simultaneously on ordinary, non-relativistic
 quantum mechanics.

 A conspicuous hallmark of extra dimensions is the appearance of
 central charges in the symmetry algebras of physical systems.
 In the context of string theory, and its effective description
 in terms of supergravity theories, these typically appear
 as central terms in superalgebras.
 Although supersymmetry central charges are a relatively mature subject in
 higher-dimensional field theories \cite{wittenolive, deWit}, relatively little
 attention has been
 applied to basic questions regarding similar charges in supersymmetric
 quantum mechanics.  Accordingly, we undertook the seemingly
 academic exercise of re-visiting the systematic development of supersymmetric
 quantum mechanics \cite{wittensqm}, with a specific intent to methodically
 build-in a non-trivial central charge.

 In this paper we critically examine the algebraic constraints that limit
 the inclusion of central terms into quantum $d=1$ superalgebras.
 We explain in detail how non-relativistic particle models
 based on supersymmetric sigma models can be extended to admit
 a non-trivial vector as a background field, in such a way that
 this vector appears as a central charge in the corresponding
 superalgebra.  We show how this can be done only if the the background
 geometry has an isometry, in which case the central charge vector
 must be a Killing vector.  We explicitly
 quantize two classes of models that conform to these
 constraints, namely models constructed on a target-space
 with topology $\R\times (S^1)^{D-1}$ and others with
 topology $\R\times T^2$.  In the second class of models, we
 demonstrate the invariance of the quantum theory under
 $SL(2\,,\,\Z)$ modular transformations which preserve the size of
 the $T^2$ factor.  In both cases we prove the existence of
 a $\Z_2$ duality which equates models with ``large" compact space
 with ostensibly distinct models having ``small" compact spaces.

 We formulate supersymmetric quantum mechanics by canonically
 quantizing a classical field theory describing the non-relativistic
 ``world-line" description of a point particle propagating in a $D$-dimensional
 Euclidean target space.
 We allow one or more of the target space dimensions to be compact.
 In the interest of simplicity, we do not in this paper
 include a superpotential {\it per se}. Instead, all interactions
 are inherited from the background geometry. The fermionic operators
 transform
 non-trivially under ``spin" transformations inherited from the
 structure group on the target space. If the central charge vanishes, then
 the quantum supercharge organizes as $Q=i\,\cDslash$,
 where ${\cal D}_m$ is a spin-covariant derivative.
 Furthermore, Kaluza-Klein interactions appear, owing to the connection pieces in
 this derivative.

 Suppose, for introductory purposes, that we have
 exactly one non-compact dimension, parameterized by $X^1$, and exactly one
 circular compact dimension parameterized by an angle $X^2$.
 Assume that the circular dimension has radius $R(X^1)$, which can depend on $X^1$.
 The fermionic operators are described by elements of a complex Clifford
 algebra,
 with elements $\Gamma^{1,2}$ and $\Gamma^{1,2\,\dagger}$, subject to
 $\{\,\Gamma^M\,,\,\Gamma^{N\,\dagger}\,\}=\d^{MN}$
 and $\{\,\Gamma^M\,,\,\Gamma^N\,\}=0$,
 where $M$ and $N$ are local frame indices.
 The complexification is required to accommodate the central
 charge; in ordinary supersymmetric quantum mechanics
 a real Clifford algebra is sufficient.
 If we systematically include a central term in the
 superalgebra in as minimal a fashion as possible
 the modified supercharge turns out to be
 \brr Q=i\,\cDslash+\mu\,R(X^1)\,\Gamma^{2\,\dagger} \,,
 \err
 where $\mu$ is a parameter associated with the central charge,
 and the slash denotes contraction with $\Gamma^M$, not with
 $\Gamma^{M\,\dagger}$. As a result, $Q$ transforms in a reducible
 spinor representation of the structure group $SO(D)$, rather than
 as an irreducible spinor.
 We can make explicit the dependence of $\cDslash$ on the angular momentum
 $i\,\der_2\equiv \nu\in\Z$, which is quantized since $X^2$ is an angular
 variable, by writing
 \brr Q=i\,\tilde{\cDslash}
      +\frac{\nu}{R(X^1)}\,\Gamma^2
      +\mu\,R(X^1)\,\Gamma^{2\,\dagger} \,.
 \label{iq}\err
 Here the operator $\tilde{\cDslash}$ includes all of the terms in
 $\cDslash$ which do not depend on $\nu$.  By writing the supercharge as in
 (\ref{iq}), one notices an amusing feature.
 Namely, the Hamiltonian, defined via
 $H=\ft12\,\{\,Q\,,\,Q^\dagger\,\}$, exhibits a duality under the
 following transformation
 \brr R(X^1) &\to& \frac{1}{R(X^1)}
      \nonumber\\[.1in]
      \mu &\leftrightarrow& \nu \,.
 \label{simt}\err
 In particular, under (\ref{simt}), the Hamiltonian undergoes a unitary
 transformation $H\to\Omega^\dagger\,H\,\Omega$, where $\Omega$
 squares to the identity.
 Thus, this class of models exhibits a
 $T$-duality, wherein models constructed with a small compact dimensions are
 physically identical to ostensibly distinct models formulated on a relatively large
 compact dimension
 \footnote{This observation was made previously by us in
 \cite{fs2}.  In that paper, however, a particular choice of fermion representation
 was imposed, so that the geometrical significance of the result was
 somewhat obscured.}.
 This scenario represents the simplest example of a
 phenomenon which appears generically in supersymmetric quantum
 mechanics when a central charge is switched on.

 Some of the discussion in this paper parallels similar
 arguments known previously in string theory.  Indeed, the dualities
 which we describe are probably closely related to string theory
 target-space dualities \cite{alvarez}.  However, we believe
 that making firm connections between the string theory phenomenon
 and the point particle analog is not a trivial exercise, and may
 include physically relevant subtlety.
 At the same time, we find it interesting how the existence of
 target space dualities can be inferred, on basic grounds, using
 modestly minimalist modification to ordinary quantum mechanics.
 We find this point of view potentially useful
 for identifying points of departure from string theory
 or for ways to connect string theory with other ideas, such as shape invariance
 or loop quantum gravity.  Indeed, owing to a conjectured relationship
 between string theory and loop quantum gravity \cite{smolin}, it seems that
 basic quantum mechanics is a natural realm to look for points
 of connection.

 Our motivation for studying centrally extended $d=1$ superalgebras
 stemmed originally from our efforts to understand the deceptively simple
 algebraic structure of shape invariance \cite{fs1}, found in
 ordinary quantum mechanics.
 Although shape invariance is not crucial to the results described
 in this paper, we feel that it is useful to mention this
 concept at the outset, since it has been an important motivator,
 and because we believe there may ultimately be some signficant
 connections between shape invariance and the work in this paper.
 We find it compelling that
 centrally extended $d=1$ superalgebras appear naturally in
 a context which has no {\it a priori} relationship to
 higher-dimensional quantum field theories.

 This paper is organized as follows.

 In section 2 we define the algebraic basis for including central terms
 into the $d=1$ $N=1$
 superalgebra.  We use superspace techniques to determine the
 transformation rules for the unique multiplet that includes
 a real commuting field as lowest component, and identify the
 modifications required to
 switch on a non-trivial central charge.  We show that the central charge
 can be incorporated as an arbitrary background vector field on
 the target space.

 In section 3 we use superspace techniques to systematically derive an
 action which is invariant under the modified transformation rules
 derived in section 2.  This action incorporates the extended real
 multiplets as fundamental fields,
 and describes a supersymmetric sigma model with a
 target-space metric as a background field.  We explain how
 this is possible only if the background central charge vector
 field and the background metric field are constrained to obey
 a system of coupled differential equations.  In this way, we show
 how the background geometry is limited by the requirement of the
 supersymmetry central charge.  We describe a class of solutions
 to this constraint.

 In section 4 we analyze a subset of the sigma models derived
 in section 3 corresponding to a class of toroidal
 compactification schemes in which the lattice describing
 the compact space is orthogonal.  We quantize this construction
 and show how the supercharge organizes to transform as a target
 space spinor, in such a way that the target space duality structure
 is manifest.

 In section 5 we analyze a class of centrally extended sigma
 models constructed on target-spaces having topology $\R\times
 T^2$.  In this case we allow an arbitrary constant complex modulus on the
 $T^2$ factor and also allow a scale factor which can depend on
 the coordinate of the non-compact dimension.  We quantize this
 model and show how the quantum supercharge organizes into a
 target-space spinor, the structure of which makes clear the existence of a
 generalization of the duality explained in section 4.

 In section 6 we study the behavior under scale-preserving
 modular transformations of the quantum supercharge
 obtained in the context of the $\R\times T^2$ compactifications
 described in section 5.  We demonstrate that the states
 in this model exhibit an appropriate $SL(2\,,\,\Z)$ symmetry
 structure so as to ensure that the scale-preserving modular
 transformations represent a symmetry.  This provides a useful
 consistency check.

 In section 7 we study the behavior under $\Z_2$ transformations
 that change the scale of the $T^2$ factor in the $\R\times T^2$
 compactifications described in section 5.  By finding an
 appropriate $\Z_2$ generator which acts on the states of the
 model, we show how these transformations describe an interesting
 generalization of the duality described in section
 4, as anticipated by the discussion in section 5.

 We conclude by making some comments on possible relationships
 between the results of this paper with other ideas, including shape
 invariance and string theory.

 \setcounter{equation}{0}
 \section{The Centrally Extended Superalgebra}
 A supercharge $Q$ is, by definition, an operator that obeys
 $\{\,Q\,,\,Q^\dagger\,\}=2\,H$, where $H$ is the Hamiltonian.
 Ordinary supersymmetric quantum
 mechanics follows from including such operators, subject to the
 additional requirement that $Q^2=0$, into the
 fundamental symmetry algebra of a physical system.
 We are interested in extending this algebra by introducing an
 additional non-trivial central charge $Z$, such that $Q^2=Z$,
 and asking what sorts of basic physics follows from this.
 Thus, we are interested in the centrally extended superalgebra
 described by
 \brr \{\,Q\,,\,Q^\dagger\,\}=2\,H
      \hspace{.3in}
      Q^2=Z
      \hspace{.3in}
      [\,Q\,,\,H\,]=0 \,.
 \label{alg}\err
 It follows trivially that $[\,Q\,,\,Z\,]=0$.  One
 also computes
 \brr [\,Q^\dagger\,,\,Z\,]
      &=& [\,Q^\dagger\,,\,Q^2\,]
      \nonumber\\[.1in]
      &=& \{\,Q^\dagger\,,\,Q\,\}\,Q
      -Q\,\{\,Q^\dagger\,,\,Q\,\}
      \nonumber\\[.1in]
      &=& 2\,[\,H\,,\,Q\,]
      \nonumber\\[.1in]
      &=& 0 \,,
 \err
 where we pass to the final line using the third
 relationship in (\ref{alg}).
 Represent the Hamiltonian as $H=i\,\der_t$, where
 $t$ is a ``time" coordinate, and represent the central charge
 by writing $Z=i\,\d_Z$ and
 $Z^\dagger=i\,\d_{Z^\dagger}$, where $\d_Z$ describes a corresponding
 central charge transformation.  Define a
 supersymmetry transformation as
 $\d_Q(\e)=\e\,Q+\e^\dagger\,Q^\dagger$,
 where $\e$ is a complex anti-commuting parameter.
 Using this, we derive
 \brr [\,\d_Q(\e_1)\,,\,\d_Q(\e_2)\,]
      &=& -4\,i\,\e_{[1}^\dagger\e_{2]}\,\der_t
      -2\,i\,\e_1\,\e_2\,\d_Z
      -2\,i\,\e_1^\dagger\,\e_2^\dagger\,\d_{Z^\dagger}
      \nonumber\\[.1in]
      [\,\d_Q(\e)\,,\,\d_Z\,] &=& 0
      \nonumber\\[.1in]
      [\,\d_Q(\e)\,,\,\d_{Z^\dagger}\,] &=& 0 \,.
 \err
 It is straightforward to find multiplet structures which represent
 these relationships.  There are a variety of possibilities.  Two
 of these are analogs of the vector multiplets and chiral
 multiplets familiar from supersymmetric field theories.  There
 also exist related multiplets with the positions of the commuting
 and anti-commuting fields in the superfield swapped, which we refer to as ``flipped"
 multiplets
 \footnote{The flipped multiplets were discussed originally in \cite{GatesKetov}
 where the operation of interchanging commuting fields and anti-commuting
 fields was referred to as a Klein flip.}.
 In this paper we keep things simple by focussing exclusively on
 real commuting multiplets.

 \subsection{Real Multiplets and Harmonic Supercharges}
 Construct a $d=1$ $N=1$ superspace by combining our real commuting ``time" coordinate
 $t$ with one additional complex anti-commuting coordinate $\t$.  Introduce
 superspace operators
 \brr Q &=& \frac{\der}{\der\,\t}
      -i\,\t^\dagger\,\der_t
      -i\,\t\,\d_Z
      \nonumber\\[.1in]
      Q^\dagger &=& \frac{\der}{\der\,\t^\dagger}
      -i\,\t\,\der_t
      -i\,\t^\dagger\,\d_{Z^\dagger} \,,
 \label{scharges}\err
 which by-construction satisfy the algebra (\ref{alg}) with the signs on
 $H$ and $Z$ reversed.  The sign-reversal is necessary, since the
 superspace coordinates $\t$ are anti-commuting, so that the
 algebra generated on superfield components by these operators
 respects (\ref{alg}).
 The inclusion of a central charge transformation
 is reminiscent of a technique used in
 supersymmetric field theories in the context of so-called
 Harmonic superspace
 \cite{harmonicss}.  Accordingly, we refer to the operators
 in (\ref{scharges}) as harmonic supercharges.
 Introduce a set of $D$ real superfields
 \brr V^n=X^n+i\,\t\,\psi^n
      +i\,\t^\dagger\,\psi^{n\,\dagger}
      +\t^\dagger\,\t\,B^n \,,
 \label{vdef}\err
 where $n=1,...,D$.  We interpret the lowest components
 $X^n$ as the spatial coordinates on a $D$-dimensional Euclidean target-space
 in which a particle, whose physics we wish to study, will propagate.
 Parameterize the particle trajectory in this space using the time
 coordinate $t$.  The superfields $V^n$ are, therefore, functions of $t$.
 A world-line supersymmetry transformation is given by
 $\d_Q(\e)=\e\,Q+\e^\dagger\,Q^\dagger$, where $\e$ is a complex
 anti-commuting parameter.  Applying
 (\ref{scharges}) to (\ref{vdef}), we derive
 \brr \d_Q(\e)\,X^n &=& i\,\e\,\psi^n
      +i\,\e^\dagger\,\psi^{n\,\dagger}
      \nonumber\\[.1in]
      \d_Q(\e)\,\psi^n &=&
      \e^\dagger\,(\,\dot{X}^n+i\,B^n\,)
      +\e\,\d_Z\,X^n
      \nonumber\\[.1in]
      \d_Q(\e)\,\psi^{n\,\dagger} &=&
      \e\,(\,\dot{X}^n-i\,B^n\,)
      +\e^\dagger\,\d_{Z^\dagger}\,X^n
      \nonumber\\[.1in]
      \d_Q(\e)\,B^n &=& \e\,\dot{\psi}^n
      -\e^\dagger\,\dot{\psi}^{n\,\dagger}
      -\e\,\d_Z\,\psi^{n\,\dagger}
      +\e^\dagger\,\d_{Z^\dagger}\,\psi^n \,.
 \label{t1}\err
 An important question is how $\d_Z$ and $\d_{Z^\dagger}$ act
 on the component fields $X^n$, $\psi^n$ and $B^n$.
 The central charge transformation should commute with complex conjugation.
 Since $X^n$ is real, this imposes $\d_Z\,X^n=\d_{Z^\dagger}\,X^n$.
 We need other commutators
 in the algebra to resolve the consistent possibilities,
 subject to this constraint. In the
 simplest class of possibilities, $\d_Z\,X^n$ appears as an
 arbitrary function of the bosonic fields,
 \brr \d_Z\,X^n &=& f^n(X)\,,
 \label{dt1}\err
 where $f^n(X)$ is an unspecified real-valued
 function of $X^1,...,X^D$.
 It is also possible to include fermion bilinears in
 $\d_Z\,X^n$.  For instance, we could write $\d_Z\,X^n=f^n(X)
 +c_{mn}(X)\,\psi^{m\,\dagger}\,\psi^n$, where $c_{mn}(X)$ is an
 unspecified real-valued symmetric tensor.
 There are several other ways in which (\ref{dt1}) could
 also be modified. However, restricting $\d_Z\,X^n$
 to depend only on $X^1,...,X^D$ provides for
 a tractable and elegant multiplet structure which admits an
 interesting class of invariant actions.
 Thus, in the spirit of minimalism,
 we restrict attention to the possibility
 described by (\ref{dt1})
 \footnote{In this paper we construct supersymmetric sigma models
 which have only a target space metric
 as a background field.  This proves possible
 given the minimal choice given in (\ref{dt1}).  In planned extensions to
 this work we intend to include additional background fields, such
 as an antisymmetric tensor. It might be necessary in such cases to
 include fermions in the transformation $\d_Z\,X^n$. }. In this case, imposing
 $[\,\d_Q\,,\,\d_Z\,]=[\,\d_Q\,,\,\d_{Z^\dagger}\,]=0$,
 requires
 \brr \d_Z\,\psi^n &=& \der_m\,f^n(X)\,\psi^m
      \nonumber\\[.1in]
      \d_Z\,\psi^{n\,\dagger} &=&
      \der_m\,f^n(X)\,\psi^{m\,\dagger}
      \nonumber\\[.1in]
      \d_Z\,B^n &=& \der_m\,f^n(X)\,B^m
      +\der_m\,\der_l\,f^n(X)\,\psi^{m\,\dagger}\,\psi^l \,,
 \label{dt23}\err
 where $\der_m=\der/\der\,X^m$.
 One derives (\ref{dt23}) by using $[\,\d_Q\,,\,\d_Z\,]=0$ together with
 (\ref{dt1}) and (\ref{t1}).  Together, these imply that the superfields transform
 as
 \brr \d_Z\,V^n &=& f^n(V) \,.
 \err
 In order that the central charge preserve the reality constraint
 $V=V^\dagger$, we also require $\d_Z=\d_{Z^\dagger}$.
 Using (\ref{dt1}) and (\ref{dt23}), the component transformation rules
 (\ref{t1}) are
 \brr \d_Q(\e)\,X^n &=& i\,\e\,\psi^n
      +i\,\e^\dagger\,\psi^{n\,\dagger}
      \nonumber\\[.1in]
      \d_Q(\e)\,\psi^n &=&
      \e^\dagger\,(\,\dot{X}^n+i\,B^n\,)
      +\e\,f^n(X)
      \nonumber\\[.1in]
      \d_Q(\e)\,\psi^{n\,\dagger} &=&
      \e\,(\,\dot{X}^n-i\,B^n\,)
      +\e^\dagger\,f^n(X)
      \nonumber\\[.1in]
      \d_Q(\e)\,B^n &=& \e\,\dot{\psi}^n
      -\e^\dagger\,\dot{\psi}^{n\,\dagger}
      -\e\,\der_m\,f^n(X)\,\psi^{m\,\dagger}
      +\e^\dagger\,\der_m\,f^n(X)\,\psi^m \,.
 \label{trans}\err
 In the case where $f^n(X)=0$ these correspond to
 the transformation rules for real supermultiplets in
 ordinary supersymmetric quantum mechanics.  The terms
 involving $f^n(X)$ describe the
 basic modifications which switch on the central charge.

 For the purpose of forming a representation of the superalgebra,
 the central charge functions $f^n(X)$ can be chosen freely;
 i.e., the representation of $Z$ is relatively unconstrained
 by the algebra.
 However, interesting restrictions on the possible choices for
 $f^n(X)$ appear if one imposes additional requirements based on
 physics, such as the existence of an invariant action functional
 involving only a finite number of time derivatives.
 In the context of supersymmetric sigma models, the possible
 choices for the functions $f^n(X)$ are correlated with the
 possible choices of sigma model metric
 \footnote{The technology described in this and in the following section
 resembles similar technology used in a two-dimensional context
 in \cite{GatesNonLinear}.  At the classical level, the
 constructions in that paper are probably related to ours by
 dimensional reduction.
 The techniques described here also usefully generalize
 some related techniques described in
 \cite{GatesSqm}, which describes the rudiments of a
 theory of linear representations of $d=1$ supersymmetry
 without central charges.}.

 \setcounter{equation}{0}
 \section{Invariant Action}
 \label{actionsec}
 In this section we
 construct invariant actions that incorporate the
 centrally extended real multiplets derived above as fundamental
 fields.  A logical method is to start with a ``lowest-order" functional
 $S_0$ whose supersymmetry variation
 vanishes when the functions $f^n(X)$ vanish.  This is easily
 accomplished by writing $S_0$ as an ordinary superspace integral.
 As minimalists, we disallow terms in the component
 Lagrangian involving more than two time derivatives.
 We also restrict attention to supersymmetric sigma models which
 include only a target space metric $g_{mn}(V)$ as a background field.
 Accordingly, we choose as a ``lowest order" action,
 \brr S_0=\ft12\,\int dt\,d\t\,d\t^\dagger\,g_{mn}(V)\,D^\dagger
      V^m\,D V^n \,,
 \label{S0}\err
 where $ds^2=g_{mn}(X)\,dX^m\,dX^n$
 describes a line element on the target space and $D$ is a superspace
 derivative
 \footnote{We have used the symbol $D$ for the target space dimensionality
 and also for the superspace derivative.  This should not cause any
 confusion, since the distinction is naturally clear from the
 context in which this symbol is used. See Appendix \ref{conventions} for a more detailed
 description of the superspace conventions and techniques employed
 in this section.}, defined as
 $D=\der/\der\t+i\,\t^\dagger\,\der_t$.
 Note that $S_0$ is
 supersymmetric in the case where $f^n(X)=0$, but
 requires modifications to restore supersymmetry when
 $f^n(X)\ne 0$ \footnote{The reason for this is
 the following. The superspace integrand in (\ref{S0}) is
 itself a real superfield.
 The highest component of $\d_Q\,V$, where $V$ is a real
 superfield, is a total derivative when $f^n(X)=0$.
 However, $\d_Q\,V$ also has terms proportional to $f^n(X)$
 which do not describe a total derivative. }.
 To systematize the analysis,
 it is useful to separate the terms in the transformation rules
 (\ref{trans}) into those terms not involving $f^n(X)$
 and those which do include these modifications.
 Accordingly, we write
 \brr \d_Q(\e)=\d_Q^{(0)}(\e)+\d_Q^{(1)}(\e) \,,
 \err
 where $\d_Q^{(0)}(\e)$ describes all terms in the transformation
 rules (\ref{trans}) which do not include $f^n(X)$.  If follows that
 $\d_Q^{(1)}(\e)$ includes all terms in (\ref{trans}) which do include these
 functions, whereby
 \brr \d_Q^{(1)}(\e)\,X^n &=& 0
      \nonumber\\[.1in]
      \d_Q^{(1)}(\e)\,\psi^n &=& \e\,f^n(X)
      \nonumber\\[.1in]
      \d_Q^{(1)}(\e)\,\psi^{n\,\dagger} &=& \e^\dagger\,f^n(X)
      \nonumber\\[.1in]
      \d_Q^{(1)}(\e)\,B^n &=&
      -\e\,\der_m\,f^n(X)\,\psi^{m\,\dagger}
      +\e^\dagger\,\der_m\,f^n(X)\,\psi^m \,.
 \label{compost}\err
 These component transformation rules (\ref{compost}) are concisely described by the following
 superfield transformation,
 \brr \d_Q^{(1)}\,V^n=
      \bpl\,i\,\t\,\e
      +i\,\t^\dagger\,\e^\dagger\,\bpr\,f^n(V) \,.
 \label{dzsup}\err
 Using the superspace variation (\ref{dzsup}), it is
 straightforward to compute the supersymmetry variation of
 (\ref{S0}).  We find
 \brr \d_Q^{(1)}\,S_0 &=&
      \ft12\,\int dt\,d\t\,d\t^\dagger\,\bl\,
      \der_l\,g_{mn}(V)\,\bpl\,i\,\t\,\e\,f^l(V)
      +i\,\t^\dagger\,\e^\dagger\,f^l(V)\,\bpr\,
      D^\dagger V^m\,D V^n
      \nonumber\\[.1in]
      && \hspace{1in}
      +g_{mn}(V)\,D^\dagger\,\bpl\,i\,\t\,\e\,f^m(V)
      +i\,\t^\dagger\,\e^\dagger\,f^m(V)\,\bpr\,D\,V^n
      \nonumber\\[.1in]
      && \hspace{1in}
      +g_{mn}(V)\,D^\dagger V^m\,D\,\bpl\,i\,\t\,\e\,f^n(V)
      +i\,\t^\dagger\,\e^\dagger\,f^n(V)\,\bpr\,\br \,.
 \label{dq1}\err
 In the case where the central charge functions $f^n(X)$ vanish,
 we see, naturally, that $S_0$ is supersymmetric, i.e.,
 $\d_Q^{(1)}\,S_0$ vanishes. In the case where
 $f^n(X)$ is non-vanishing, $S_0$ ceases to be supersymmetric by itself.
 To restore supersymmetry, we therefore must add to $S_0$ new
 terms whose supersymmetry variation cancels against
 (\ref{dq1}).

 This process is systematized by the following sequence of operations.
 First, if possible, construct a superspace functional $S_1$
 with the property
 $\d_Q^{(0)}\,S_1=-\d_Q^{(1)}\,S_0$.  The supersymmetry
 variation of the sum $S_0+S_1$ is then given by
 $\d_Q^{(1)}\,S_1$.  If this is non-vanishing, then iterate
 this procedure by constructing another superspace functional
 $S_2$ with the property $\d_Q^{(0)}\,S_2=-\d_Q^{(1)}\,S_1$.
 As we will show with explicit calculation, in those
 cases where one can construct $S_1$ and $S_2$ according to the
 above prescription, the superspace integrand
 in $S_2$ turns out to be quadratic in fermionic coordinates, i.e., this
 expression is proportional to $\t^\dagger\,\t$.  Since the
 operator (\ref{dzsup}) is itself linear in $\t$ and $\t^\dagger$,
 it follows that $\d_Q^{(1)}\,S_2=0$.  Therefore, this procedure
 terminates after two iterations, and the combination
 $S_0+S_1+S_2$ is supersymmetric.  An important question
 remains: under what circumstances can one construct
 $S_1$ and $S_2$ according to our prescription?

 It is useful to re-write equation (\ref{dq1})
 in a more useful form. After a small amount of algebra, one finds
 \brr \d_Q^{(1)}\,S_0 &=&
      \int dt\,d\t\,d\t^\dagger\,\bl\,
      \ft12\,g_{mn}(V)\,
      \bpl\,i\,\e^\dagger\,f^m(V)\,D\,V^n
      -i\,\e\,f^m(V)\,D^\dagger V^n\,\bpr
      \nonumber\\[.1in]
      & & \hspace{1in}
      +\bpl\,i\,\t\,\e+i\,\t^\dagger\,\e^\dagger\,\bpr\,
      \Omega_{mn}(V)\,
      D^\dagger V^m\,D\,V^n\,\br \,,
 \label{vartwo}\err
 where
 \brr \Omega_{mn}(V)= g_{l(m}(V)\,\der_{n)}\,f^l(V)
      +\ft12\,f^l(V)\,\der_l\,g_{mn}(V) \,.
 \err
 Using the definition of the affine connection,
 $\Gamma_{mn}\,^l=\ft12\,g^{lr}\,(\,\der_m\,g_{nr}+\der_n\,g_{mr}-\der_r\,g_{mn}\,)$,
 it is straightforward to prove that
 $\Omega_{mn}=\nabla_{(m}\,f_{n)}$,
 where $\nabla_m\,f_n=\der_m\,f_n-\Gamma_{mn}\,^l\,f_l$ is a
 derivative covariant
 with respect to target space coordinate transformations.

 The second line in (\ref{vartwo}) has the following special
 feature.  If we replace $\e$ with $\t$ and replace $\e^\dagger$
 with $\t^\dagger$, then this line vanishes identically.  As explained in
 detail in Appendix \ref{conventions}, this structure tells us
 that this line cannot represent a basic supersymmetry variation; that
 is, this line does not represent $\d_Q^{(0)}$ of any expression.
 Therefore, our only hope for finding a supersymmetric
 extension to $S_0$ is if this line vanishes identically.  Accordingly,
 we must insist that the target space metric components and
 the central charge functions are correlated in such a way
 that $\nabla_{(m}\,f_{n)}$ vanishes.  This implies
 that the system of coupled differential equations
 defined by $\nabla_{(m}\,f_{n)}=0$ is satisfied.
 There is another way to understand this condition.
 Notice that the transformation of $S_0$ under the central charge
 is
 \brr \d_Z\,S_0 &=&
      \int dt\,d\t\,d\t^\dagger\,
      \nabla_{(m}\,f_{n)}(V)\,D^\dagger V^m\,D\,V^n \,.
 \err
 Thus, the requirement $\d_Z\,S_0=0$ is equivalent to the
 requirement that we can find proper counter-terms $S_1$ to cancel
 $\d_Q\,S_0$.  When we impose the condition $\nabla_{(m}\,f_{n)}=0$,
 equation (\ref{vartwo}), simplifies to
 \brr \d_Q^{(1)}\,S_0 &=&
      \ft12\,\int dt\,d\t\,d\t^\dagger\,
      g_{mn}(V)\,
      \bpl\,i\,\e^\dagger\,f^m(V)\,D\,V^n
      -i\,\e\,f^m(V)\,D^\dagger V^n\,\bpr \,.
 \label{dqa}\err
 Now, following our procedure, we need to find an $S_1$
 which has the property
 $\d_Q^{(0)}\,S_1= -\d_Q^{(1)}\,S_0$.  This is achieved by
 \brr S_1 &=&
      \ft12\,\int dt\,d\t\,d\t^\dagger\,
      g_{mn}(V)\,
      \bpl\,i\,\t^\dagger\,f^m(V)\,D\,V^n
      -i\,\t\,f^m(V)\,D^\dagger V^n\,\bpr \,.
 \label{s1zz}\err
 To obtain this, we simply replace each instance of $\e$ in
 (\ref{dqa}) with $\t$ and each instance of $\e^\dagger$ with
 $\t^\dagger$.
 Now consider the next order in the supersymmetry variation.
 After some algebra, we derive
 \brr \d_Q^{(1)}\,S_1 &=&
      -\int dt\,d\t\,d\t^\dagger\,\bl\,
      \ft12\,\bpl\,\t\,\e^\dagger-\t^\dagger\,\e\,\bpr\,
      g_{mn}(V)\,f^m(V)\,f^n(V)
      \nonumber\\[.1in]
      & & \hspace{1in}
      +i\,\t^\dagger\,\t\,\nabla_{(m}\,f_{n)}(V)\,f^m(V)\,
      \bpl\,\e\,D\,V^n+\e^\dagger\,D^\dagger V^n\,\bpr\,\br \,.
 \label{tutu}\err
 Since $\nabla_{(m}\,f_{n)}$ vanishes for any of the
 allowable backgrounds, equation (\ref{tutu}) automatically simplifies to
 \brr \d_Q^{(1)}\,S_1 &=&
      -\ft12\,\int dt\,d\t\,d\t^\dagger\,\bpl\,
      \t\,\e^\dagger-\t^\dagger\,\e\,\bpr\,
      g_{mn}(V)\,f^m(V)\,f^n(V) \,.
 \label{d1zz}\err
 The variation (\ref{d1zz}) is cancelled by adding terms $S_2$
 having the property
 $\d_Q^{(0)}\,S_2=-\d_Q^{(1)}\,S_1$.  This is achieved by
 \brr S_2=-\ft12\,\int dt\,d\t\,d\t^\dagger\,
      \t^\dagger\,\t\,g_{mn}(V)\,f^m(V)\,f^n(V)  \,.
 \label{s2zz}\err
 To obtain this, replace each instance of $\e$ in
 (\ref{d1zz}) with $\t$ and each instance of $\e^\dagger$ with
 $\t^\dagger$, and divide by two, since the ultimate result is quadratic
 in $\t$ and $\t^\dagger$.  We see that $\d_Q^{(1)}\,S_2=0$, so that the
 sum $S=S_0+S_1+S_2$ is supersymmetric.  Adding up the terms
 (\ref{S0}), (\ref{s1zz}) and (\ref{s2zz}), and then factorizing, we obtain
 \brr S &=& \ft12\,\int dt\,d\t\,d\t^\dagger\,
      g_{mn}(V)\,\bpl\,D^\dagger V^m+i\,\t^\dagger\,f^m(V)\,\bpr\,
      \bpl\,D\,V^n+i\,\t\,f^n(V)\,\bpr
 \label{key}\err
 where the metric $g_{mn}(V)$ and the central charge functions
 $f^m(V)$ are constrained by
 \brr  \nabla_{(m}\,f_{n)}=0 \,.
 \label{backdif}\err
 Notice that this is Killing's equation.  Thus, the allowed
 central charge functions $f^n(X)$ must organize as the components
 of a Killing vector.  This tells us that the background geometry
 must possess an isometry in order for the sigma model to admit
 a supersymmetry central charge.

 Our goal in this paper is to address the basic
 features of interest that appear when the supersymmetry
 central charges are switched on. Therefore, rather than describe general
 solutions to (\ref{backdif}), we restrict attention to
 the simplest class of allowable target space metrics
 that exhibit novel features related to the central extension.
 We plan to address more general backgrounds in
 more comprehensive future work.

 \subsection{Consistent Backgrounds}
 A simple way to satisfy (\ref{backdif}) is to
 consider a target space manifold with topology
 \brr {\cal M}^D={\cal S}^p\times X^{D-p} \,,
 \label{top1}\err
 where ${\cal S}^p$ is a $p$-dimensional ``space" and
 $X^{D-p}$ is a $(D-p)$-dimensional ``internal" space.  The metric
 decomposes as $g=g_{\cal S}\otimes g_X$ where $g_{\cal S}$ is the metric on
 ${\cal S}^p$, which we do not let depend on the coordinates on $X^{D-p}$,
 and $g_X$ is the metric on the internal space.
 The internal metric $g_X$ may, in general, depend on any of the
 $D$ coordinates on ${\cal M}^D$.  The coordinates on ${\cal M}^D$ are
 given by the bosonic component fields $X^1,...,X^D$.
 Correspondingly, the superscript $n$ which appears
 on the superfields $V^n$ and on the central charge functions $f^n(V)$ describes
 a contravariant target space vector index.  The central charge
 functions then describe a vector which decomposes as
 $f=f_{\cal S}\otimes f_X$,
 where $f_{\cal S}\equiv(f^1,...,f^p)$ is a vector on ${\cal S}$ and
 $f_X\equiv(f^{D+1},...,f^d)$ is a vector on the internal space $X$.
 If we choose $f_{\cal S}=0$ and restrict the metric to depend only
 the coordinates on ${\cal S}^p$; i.e., $g=g(\,X^1,...,X^p\,)$,
 then (\ref{backdif}) is satisfied.

 In this paper we not only restrict our attention to the manifolds
 described in the previous paragraph, but we further
 simplify to a case involving a flat target space
 \brr {\cal M}^D=\R^p\times X^{D-p}
 \err
 where $X^{D-p}$ is a $(D-p)$-dimensional torus.
 This is done in the interest of stripping down
 the basic physics implied by supersymmetry central charges to its
 essence. In future work we intend to study the extra
 ramifications which follow from more general choices in the
 class of allowable backgrounds.

 \setcounter{equation}{0}
 \section{A class of toroidal compactifications}
 \label{torrid}
 It is instructive to specialize to the following case.
 Restrict the target space to have topology $\R\times (\,S^1\,)^{D-1}$.  Let
 $X^1\in\R$
 parameterize the non-compact dimension, and let
 $X^{i\ne 1}\in [\,0\,,\,2\,\pi\,]$ describe one angular coordinate on
 each compact dimension.
 The compact dimensions are taken as circles having radii
 $R_i(X^1)$, which can depend independently on $X^1$.  Accordingly,
 choose the metric
 \brr ds^2=(\,dX^1\,)^2+\sum_{i=2}^D\,R_i(X^1)^2\,(\,dX^i\,)^2\,.
 \label{metric}\err
 By convention, indices $i,j,k$ enumerate
 compact dimensions, whereas indices $m,n,p$ enumerate
 all dimensions.  Thus, $i=2,...,D$, whereas $n=1,2,...,D$.
 The class of metrics (\ref{metric}) describe a restricted class of
 toroidal compactification schemes in which the lattice describing the
 torus is orthogonal. (We generalize this to include
 a slightly more general class of lattices in the following
 section.)
 Furthermore, let the central charge functions $f^n(X)$
 be constant real numbers defined by
 \brr (\,f^1\,,\,f^2\,,\,...\,,\,f^d\,)
      &\equiv& (\,0\,,\,\mu^2\,,\,\mu^3\,,\,...\,,\,\mu^D\,) \,.
 \err
 Thus, the central charge is parameterized by one real number
 $\mu^i$ for each compact dimension.
 Following the procedure described in section \ref{actionsec},
 the action invariant under centrally extended
 supersymmetry is
 \brr S &=& \int dt\,d\t\,d\t^\dagger\,\bl\,
      \ft12\,D^\dagger\,V^1\,D\,V^1
      +\ft12\,\sum_{i=2}^D\,R_i(V^1)^2\,
      D^\dagger\,V^i\,D\,V^i
      \nonumber\\[.1in]
      & & \hspace{.9in}
      +\ft12\,\sum_{i=2}^D\,\bpl\,
      i\,\mu^i\,(\,\t^\dagger\,D-\t\,D^\dagger\,)\,V^i
      -(\,\mu^i\,)^2\,\t^\dagger\,\t\,\bpr\,\br \,.
 \label{ssac}\err
 The first line in (\ref{ssac}) describes an ordinary
 supersymmetric sigma model.  The
 second line includes terms which extend the basic
 supersymmetry so as to switch on the desired central charge.
 The component Lagrangian corresponding to (\ref{ssac}) is
 \brr L &=&
      \ft12\,\dot{X}^1\,\dot{X}^1
      -\ft12\,i\,\psi^{1\,\dagger}\derbar_t\psi^1
      +\ft12\,B^1\,B^1
      \nonumber\\[.1in]
      & & +\sum_{i=2}^D\,\bl\,
      R_i(X^1)^2\,\bpl\,
      \ft12\,\dot{X}^i\,\dot{X}^i
      -\ft12\,i\,\psi^{i\,\dagger}\derbar_t\psi^i
      +\ft12\,B^i\,B^i\,\bpr
      \nonumber\\
      & & \hspace{.5in}
      +2\,i\,R_i(X^1)\,R_i'(X^1)\,
      \psi^{[1\,^\dagger}\psi^{i]}\,\dot{X}^i
      \nonumber\\[.1in]
      & & \hspace{.5in}
      +R_i(X^1)\,R_i'(X^1)\,\bpl\,
      \psi^{1\,\dagger}\psi^i\,B^i
      +\psi^{i\,\dagger}\psi^1\,B^i
      -\psi^{i\,\dagger}\psi^i\,B^1\,\bpr
      \nonumber\\[.1in]
      & & \hspace{.5in}
      -\bpl\,R_i(X^1)\,R_i''(X^1)+R_i'(X^1)^2\,\bpr\,\psi^{i\,\dagger}\psi^i
      \psi^{1\,\dagger}\psi^1
      \nonumber\\[.1in]
      & & \hspace{.5in}
      -i\,\mu^i\,R_i(X^1)\,R_i'(X^1)\,
      \bpl\,\psi^1\,\psi^i+\psi^{1\,\dagger}\psi^{i\,\dagger}\,\bpr
      -\ft12\,(\,\mu^i\,)^2\,R_i(X^1)^2\,\br \,.
 \label{lagrangian}\err
 The action $S=\int dt\,L$ is
 invariant under the supersymmetry transformations (\ref{trans})
 and also under the $(D-1)$ independent transformations
 $\d_Z\,X^i=\mu^i$.

 The classically-conserved charges are obtained as
 follows.
 Under a supersymmetry transformation (\ref{trans}), we find
 $\d_Q\,L=\dot{K}$, where
 \brr K &=&
      \ft12\,i\,\e\,(\,\dot{X}^1-i\,B^1\,)\,\psi^1
      +\ft12\,\sum_{i=2}^D\,
      i\,\e\,R_i(X^1)\,(\,\dot{X}^i-i\,B^i\,)\,\psi^i
      \nonumber\\
      & &
      -\sum_{i=2}^D\,\bpl\,
      \e\,R_i(X^1)\,R_i'(X^1)\,\psi^{i\,\dagger}\psi^i\psi^1
      +\ft12\,i\,\e\,\mu^i\,R_i(X^1)^2\,\psi^{i\,\dagger}\,\bpr
      \nonumber\\[.1in]
      & &
      +{\rm h.c.}
 \err
 The parameter-dependent supercharge, determined by the
 Noether procedure, is given by
 \brr \tilde{Q}=
      \d_Q\,X^m\,P_m+\d_Q\,\psi^m\,\Pi_{\psi^m}
      +\d_Q\,\psi^{m\,\dagger}\,\Pi_{\psi^{m\,\dagger}}-K \,,
 \label{qnoether}\err
 where $P_m=g_{mn}(X)\,\dot{X}^n-i\,g_{m[n,p]}(X)\,\psi^{n\,\dagger}\,\psi^p$
 is the momentum conjugate to
 $X^m$, $\Pi_{\psi^m}=-\ft12\,i\,g_{mn}\,\psi^{n\,\dagger}$ is
 the momentum conjugate to $\psi^m$, and
 $\Pi_{\psi^{m\,\dagger}}=-\ft12\,i\,g_{mn}\,\psi^n$
 is the momentum conjugate to $\psi^{m\,\dagger}$.
 Now write
 $\tilde{Q}=i\,\e\,Q+i\,\e^\dagger\,Q^\dagger$,
 which defines $Q$ as the parameter-independent supercharge,
 with phase chosen as a matter of convention.
 In this way, we determine
 \brr Q &=&
      P_m\,\psi^m
      +\sum_{i=2}^D\,\bpl\,
      -i\,R_i(X^1)\,R_i'(X^1)\,\psi^{i\,\dagger}\,\psi^i\,\psi^1
      +\mu^i\,R_i(X^1)^2\,\psi^{i\,\dagger}\,\bpr \,.
 \label{noether}\err
 The conserved central charge, which is determined similarly,
 is given by
 \brr  Z &=& \sum_{i=2}^D\,\mu^i\,P_i \,.
 \label{zboo}\err
 In a similar way, one can compute the Noether Hamiltonian,
 defined as $H=\dot{X}^m\,P_m+\dot{\psi}^m\,\Pi_{\psi^m}
 +\dot{\psi}^{m\,\dagger}\,\Pi_{\psi^{m\,\dagger}}-L$.
 After some algebra, one readily verifies that the expression
 determined in this way is the same as
 $H=\ft12\,\{\,Q\,,\,Q^\dagger\,\}$.

 \subsection{Quantization}
 The quantum operator algebra, obtained from the Dirac brackets
 associated with (\ref{lagrangian}), is described by
 \brr [\,P_m\,,\,X^n\,] &=& i\,\d_m\,^n
      \nonumber\\[.1in]
      \{\,\psi^1\,,\,\psi^{1\,\dagger}\,\} &=& 1
      \nonumber\\[.1in]
      \{\,\psi^i\,,\,\psi^{j\,\dagger}\,\} &=&
      R_i^{-2}\,\d^{ij}
      \nonumber\\[.1in]
      [\,P_1\,,\,\psi^i\,] &=&
      -i\,R_i'\,R_i^{-1}\,\psi^i
      \nonumber\\[.1in]
      [\,P_1\,,\,\psi^{i\,\dagger}\,] &=&
      -i\,R_i'\,R_i^{-1}\,\psi^{i\,\dagger} \,,
 \label{canit}\err
 where $i=2,...,D$. Achieve this by writing
 $P_m=i\,\der_m$ and
 \brr \psi^1 &=& \Gamma^1
      \nonumber\\[.1in]
      \psi^i &=& \frac{1}{R_i}\,\Gamma^i \,,
 \err
 where $\Gamma^M=(\,\Gamma^1,...,\Gamma^D\,)$ are elements
 of a complex Clifford algebra\footnote{The index $M$ can be
 interpreted as a local frame index.
 More specifically, we quantize by writing $\psi^m=\Gamma^m$, where
 $\{\,\Gamma^m\,,\,\Gamma^{n\,\dagger}\,\}=g^{mn}(X)$.
 The coordinate index $m$ is replaced by a tangent
 space index $M$ by writing $\Gamma^m=\Gamma^M\,\tilde{E}_M\,^m$,
 where $\tilde{E}_M\,^m={\rm diag}(1,R_1^{-1},...,R_D^{-1})$
 is an inverse vielbein.  We have chosen a particular frame
 in writing (\ref{canit}).  As a result, the target space transformation
 properties are not manifest in many of the expressions in this and also in
 the following section.}
 $\{\,\Gamma^M\,,\,\Gamma^{N\,\dagger}\,\}=\d^{MN}$.  Since $X^i$ are
 angular coordinates, it follows that the momenta $P_i$
 are integer quantized.
 Thus, $P_i\equiv \nu_i\in\Z$.  Using these results,
 and after resolving a few ordering ambiguities,
 the quantum supercharge corresponding to (\ref{noether}) is found to be
 \brr Q &=& i\,\der_1\,\Gamma^1
      +\ft12\,i\,\sum_{i=2}^D\,\frac{R_i'}{R_i}\,
      [\,\Gamma^i\,,\,\Gamma^{i\,\dagger}\,]\,\Gamma^1
      +\sum_{i=2}^D\,\bpl\,\frac{\nu_i}{R_i}\,\Gamma^i
      +\mu^i\,R_i\,\Gamma^{i\,\dagger}\,\bpr \,.
 \label{char}\err
 Similarly, the quantum central charge is
 \brr Z &=& \sum_{i=2}^D\,\mu^i\,\nu_i \,,
 \err
 and the Hamiltonian is $H=\ft12\,\{\,Q\,,\,Q^\dagger\,\}$ \,.
 The ordering ambiguities mentioned above are found in the
 fermion cubic term in $Q$ and in the fermion quartic term
 in $H$.  After some determined algebra, one finds that
 these terms can be ordered so that $Q^2=Z$ and
 $H=\ft12\,\{\,Q\,,\,Q^\dagger\,\}$.  The result of this
 work is reflected in the particular ordering which appears in
 (\ref{char}).

 It proves illuminating to compute the
 components of the spin-connection on the target space, as
 explained in Appendix \ref{cders}.
 In doing so, one finds that the terms in (\ref{char}) that are
 cubic in the $\Gamma^M$'s organize into
 spin connection pieces which, when combined with the
 ordinary derivatives appearing in $Q$, form a spin covariant
 derivative.  In this way, one finds that the expression
 for $Q$ given in (\ref{char}) organizes as
 \brr Q=i\,\cDslash
      +\sum_{i=2}^D\, \mu^i\,R_i\,\Gamma^{i\,\dagger} \,,
 \label{limlam}\err
 where $\cDslash$ is the spin-covariant derivative
 \footnote{This is
 explained in detail in Appendix \ref{cders}, where the computation
 is done quite explicitly in the case involving one compact
 dimension.}. It is instructive to
 separate out the terms in this derivative that depend on
 $\nu_i$, by re-writing (\ref{limlam}) as
 \brr Q=i\,\tilde\cDslash
      +\sum_{i=2}^D\,\bpl\,\frac{\nu_i}{R_i}\,\Gamma_i
      +\mu_i\,R_i\,\Gamma^{i\,\dagger}\,\bpr \,,
 \err
 where $\tilde\cDslash$ is the spin covariant derivative minus all
 terms which depend on $\nu_i$.  Written this way, a certain
 duality structure becomes manifest.  Specifically,
 under the transformation
 \brr \mu^i &\leftrightarrow& \nu_i
      \nonumber\\[.1in]
      R_i &\leftrightarrow& \frac{1}{R_i} \,,
 \label{dualt}\err
 one finds $Q\to\Omega^\dagger\,Q^\dagger\,\Omega$, where
 $\Omega$ is a unitary operator which generates a $\Z_2$
 parity operation as follows,
 \brr \Omega\,\Gamma^1\,\Omega^\dagger &=& \Gamma^{1\,\dagger}\
      \nonumber\\[.1in]
      \Omega\,\Gamma^i\,\Omega^\dagger &=& \Gamma^i \,.
 \label{twine}\err
 The Hamiltonian is given by $H=\ft12\,\{\,Q\,,\,Q^\dagger\,\}$.
 Under the above transformation, we have $H\to
 \tilde{H}=\Omega^\dagger\,H\,\Omega$.  Since $H$ and $\tilde{H}$
 are related by a unitary transformation, it follows that $H$ and
 $\tilde{H}$ are iso-spectral.  Thus, the
 transformation (\ref{dualt}) represents a duality.

 \setcounter{equation}{0}
 \section{SQM on $\R\times T^2$}
 In this section we generalize the results of the previous
 section to include a twist angle into the
 internal metric corresponding to compactification on
 a two-torus. Thus, we consider a target space having topology $\R\times T^2$.
 Parameterize the noncompact dimension using $X^1\in\R$,
 and parameterize the $T^2$ factor using two angular
 variables $X^{2,3}\in [\,1\,,\,2\,\pi\,]$.
 Characterize the two-torus using modular parameter
 \brr \tau=\frac{R_3}{R_2}\,e^{i\,\a} \,,
 \err
 where $R_2$ and $R_3$ are the radii of the circles corresponding to the
 respective coordinates $X^2$ and $X^3$, and
 $\a$ is an arbitrary phase.  In this case, the target space
 metric is
 \brr ds^2 = (\,dX^1\,)^2+R_2^2\,(\,dX^2\,)^2
      +2\,R_2\,R_3\,\cos\a\,dX^2\,dX^3
      +R_3^2\,(\,dX^3\,)^2 \,.
 \label{dsm}\err
 This is the same as (\ref{metric}) in the case $D=3$ except for the new off-diagonal
 term which manifests a non-trivial twist.
 It is convenient to define a
 complex coordinate $Y = X^2+\tau\,X^3$
 and also a scale factor $\phi(X^1)$ according to
 \brr R_2(X^1) \equiv e^{\phi(X^1)} \,.
 \err
 In the case where the modulus $\tau$ does not
 depend on $X^1$, the metric (\ref{dsm}) is more
 concisely expressed as
 $ds^2 = (\,dX^1\,)^2+e^{2\,\phi(X^1)}\,|\,dY\,|^2$.  For the computational
 purposes used in this paper, we find (\ref{dsm}) more convenient,
 however.

 The  supersymmetric action is the same as that given in
 section \ref{torrid} plus new terms which correspond to the
 cross terms in the metric.  Thus, the action is given by
 $S_{\rm old}+S_{\rm new}$ where $S_{\rm old}$ is
 given in (\ref{ssac}) and
 \brr S_{\rm new}=
      {\rm Re}\,\tau\,\int dt\,d\t\,d\t^\dagger\,
      e^{2\,\phi(V^1)}\,
      \bpl\,D^\dagger V^{(2}\,D V^{3)}
      +i\,\mu^{(2}\,
      (\,\t^\dagger\,D-\t\,D^\dagger\,)\,V^{3)}
      -\mu^2\,\mu^3\,\t^\dagger\,\t\,\bpr \,.
      \nonumber\\
 \err
 The component lagrangian is $L_{\rm old}+L_{\rm new}$, where
 $L_{\rm old}$ is the lagrangian given in (\ref{lagrangian}),
 restricted to the case $D=3$, and
 \brr L_{\rm new} &=&
      2\,{\rm Re}\,\tau\,e^{2\,\phi(X^1)}\,\bl\,
      \ft12\,\bpl\,\dot{X}^2\,\dot{X}^3
      +B^2\,B^3
      -i\,\psi^{(2\,\dagger}\derbar_t\psi^{3)}\,\bpr
      \nonumber\\[.1in]
      & & \hspace{1in}
      -i\,\phi'(X^1)\,\bpl\,\dot{X}^{(2}\,\psi^{3)}\,\psi^{1\,\dagger}
      +\dot{X}^{(2}\,\psi^{3)\,\dagger}\psi^1\,\bpr
      \nonumber\\[.1in]
      & & \hspace{1in}
      +\phi'(X^1)\,\bpl\,
      B^{(2}\,\psi^{3)\,\dagger}\,\psi^1
      -B^{(2}\,\psi^{3)}\,\psi^{1\,\dagger}
      -B^1\,\psi^{(2\,\dagger}\psi^{3)}\,\bpr
      \nonumber\\[.1in]
      & & \hspace{1in}
      -\bpl\,\phi''(X^1)+3\,\phi'(X^1)^2\,\bpr\,
      \psi_1^\dagger\psi_1\,\psi_{(2}^\dagger\psi_{3)}
      \nonumber\\[.1in]
      & & \hspace{1in}
      +i\,\phi'(X^1)\,\bpl\mu^{(2}\,\psi^{3)\,\dagger}\psi^{1\,\dagger}
      +\mu^{(2}\,\psi^{3)}\psi^1\,\bpr
      -\mu^2\,\mu^3\,\br
      \,.
 \label{lnew}\err
 Since $L_{\rm new}$ is supersymmetric, it follows that the
 supersymmetric variation of (\ref{lnew}) is a total derivative,
 i.e., $\d_Q\,L_{\rm new}=\dot{K}_{\rm new}$.
 Determined calculation yields
 \brr K_{\rm new}=
      {\rm Re}\,\tau\,e^{2\,\phi(X^1)}\,\e\,\bpl\,
      i\,\dot{X}^{(2}\psi^{3)}
      +B^{(2}\,\psi^{3)}
      -2\,\phi'(X^1)\,\psi^1\,\psi^{(2\,\dagger}\psi^{3)}
      -i\,\mu^{(2}\,\psi^{3)\,\dagger}\,\bpr+{\rm h.c.}
 \err
 The ``new" contributions to the parameter-dependent supercharge
 (\ref{qnoether}) are
 \brr \tilde{Q}_{\rm new}=
      \d_Q\psi^i\,\Pi_{\psi^i}^{({\rm new})}
      +\d_Q\psi^{i\,\dagger}\,\Pi_{\psi^{i\,\dagger}}^{({\rm new})}
      -K_{\rm new} \,.
 \err
 where
 \brr \Pi_{\psi^{2,3}}^{({\rm new})}
      &=& \ft12\,i\,{\rm Re}\,\tau\,e^{2\,\phi(X^1)}\,
      \psi^{3,2\,\dagger}
      \nonumber\\[.1in]
      \Pi_{\psi^{2,3\,\dagger}}\,^{({\rm new})}
      &=& \ft12\,i\,{\rm Re}\,\tau\,e^{2\,\phi(X^1)}\,
      \psi^{3,2} \,.
 \err
 The parameter-independent supercharge $Q$ is defined via
 $\tilde{Q}=i\,\e\,Q+i\,\e^\dagger\,Q^\dagger$.
 In this way, we determine
 \brr Q_{\rm new} &=&
      2\,{\rm Re}\,\tau\,e^{2\,\phi(X^1)}\,\bpl\,
      -i\,\phi'(X^1)\,\psi_1\,\psi^{(2\,\dagger}\psi^{3)}
      +\mu^{(2}\,\psi_{3)}^\dagger\,\bpr \,.
 \err
 The full supercharge is obtained by re-writing $Q_{\rm old}$,
 given in (\ref{noether}), in
 terms of the re-defined parameters $\tau$ and $\phi(X^1)$,
 and then adding the result to $Q_{\rm new}$.
 Thus, the classically-conserved Noether supercharge is
 \brr Q &=& P_1\,\psi^1
      +P_2\,\psi^2
      +P_3\,\psi^3
      \nonumber\\[.1in]
      & &
      +e^{2\,\phi}\,\bpl\,
      -i\,\phi'\,\psi^{2\,\dagger}\,\psi^2\,\psi^1
      +\mu^2\,\psi^{2\,^\dagger}\,\bpr
      \nonumber\\[.1in]
      & &
      +e^{2\,\phi}\,|\,\tau\,|^2\,\bpl\,
      -i\,\phi'\,\psi^{3\,\dagger}\,\psi^3\,\psi^1
      +\mu^3\,\psi^{3\,\dagger}\,\bpr
      \nonumber\\[.1in]
      & &
      +2\,e^{2\,\phi}\,{\rm Re}\,\tau\,\bpl\,
      -i\,\phi'\,\psi^{(2\,\dagger}\,\psi^{3)}\psi^1
      +\mu^{(2}\,\psi^{3)\,\dagger} \,\bpr \,.
 \err
 In the quantum supercharge, the operator $P_1$ is replaced
 with $i\,\der_1$, and the fermion cubic terms organize into
 spin connection pieces which, when combined with the
 ordinary derivatives appearing in $Q$, form a spin covariant derivative
 \footnote{See Appendix \ref{cders} for details.}.  Furthermore, since $X^{2,3}$ are angular
 variables, it follows that the momenta $P_{2,3}$ are quantized as
 integers.  Thus, we write $P_{2,3}\equiv\nu_{2,3}\in\Z$.
 Accordingly, the quantum supercharge is
 \brr Q &=& i\,\tilde{\cDslash}
      +\nu_2\,\psi^2+\nu_3\,\psi^3
      \nonumber\\[.1in]
      & & +e^{2\,\phi}\,\bpl\,
      \mu^2\,\psi^{2\,\dagger}
      +|\,\tau\,|^2\,\mu^3\,\psi^{3\,\dagger}
      +2\,{\rm Re}\,\tau\,\mu^{(2}\,\psi^{3)\,\dagger}\,\bpr \,,
 \label{classq}\err
 where $\tilde\cDslash$ is the spin-covariant derivative
 minus the terms that include $\nu_{2,3}$.  This is explained
 more completely in the following subsection.

 \subsection{Quantization}
 The quantum operator algebra, obtained from the Dirac brackets,
 is described by
 \brr [\,P_m\,,\,X^n\,] &=& i\,\d_m\,^n
      \nonumber\\[.1in]
      \{\,\psi^m\,,\,\psi^{n\,\dagger}\,\} &=&
      g^{mn}(X)
      \nonumber\\[.1in]
      [\,P_m\,,\,P_n\,] &=&
      \ft14\,i\,g^{pq}\,\der_m\,g_{pq}\,\der_n\,g_{qr}\,
      \psi^q\psi^r
      \nonumber\\[.1in]
      [\,P_m\,,\,\psi^n\,] &=&
      -\ft12\,i\,g^{np}\,\der_m g_{pq}\,\psi^q \,,
 \label{quantum}\err
 where $g_{mn}(X)$ is the target space metric and $g^{mn}(X)$ is its inverse.
 This result is valid
 for any model described by (\ref{key}).  For the case at hand,
 the metric is
 \brr g_{mn} &=&
      e^{2\,\phi}\,\ba{c|cc}
      e^{-2\,\phi} &&\\
      \hline
      &1 & {\rm Re}\,\tau\\
      & {\rm Re}\,\tau & |\,\tau\,|^2 \ea\,.
 \label{taumet}\err
 This can be written in terms of a dreibein $E_m\,^M$,
 defined by $g_{mn}=E_m\,^M\,E_n\,^N\,\d_{MN}$.  We can choose
 \brr E_m\,^M=e^\phi\,\ba{c|cc}e^{-\phi}&&\\\hline
      &1&0\\&{\rm Re}\,\tau&{\rm Im}\,\tau\ea \,,
 \label{drei}\err
 in which case the inverse dreibein is
 \brr \tilde{E}_M\,^m=e^{-\phi}\ba{c|cc}
      e^\phi&&\\\hline
      &1&0\\
      & -{\rm Re}\,\tau\,(\,{\rm Im}\,\tau\,)^{-1} &
      (\,{\rm Im}\,\tau\,)^{-1}\ea \,.
 \label{inversedrei}\err
 Using the metric (\ref{taumet}), the quantum algebra (\ref{quantum})
 is given by
 \brr [\,P_m\,,\,X^n\,] &=& i\,\d_m\,^n
      \hspace{.8in}
      \{\,\psi^1\,,\,\psi^{1\,\dagger}\,\}=1
      \nonumber\\[.1in]
      [\,P_m\,,\,P_n\,] &=& 0
      \hspace{1.1in}
      \{\,\psi^2\,,\,\psi^{2\,\dagger}\,\} =
      (\,{\rm Im}\,\tau\,)^{-2}\,|\,\tau\,|^2\,e^{-2\,\phi}
      \nonumber\\[.1in]
      [\,P_1\,,\,\psi^2\,] &=& -i\,\phi'(X^1)\,\psi^2
      \hspace{.3in}
      \{\,\psi^3\,,\,\psi^{3\,\dagger}\,\} =
      (\,{\rm Im}\,\tau\,)^{-2}\,e^{-2\,\phi}
      \nonumber\\[.1in]
      [\,P_1\,,\,\psi^3\,] &=& -i\,\phi'(X^1)\,\psi^3
      \hspace{.3in}
      \{\,\psi^3\,,\,\psi^{2\,\dagger}\,\} =
      -{\rm Re}\,\tau\,(\,{\rm Im}\,\tau\,)^{-2}\,e^{-2\,\phi} \,.
 \err
 In general, we can represent (\ref{quantum})
 by writing $P_n=i\,\der_n$ and
 $\psi^m=\Gamma^M\,\tilde{E}_M\,^m$, where $\Gamma^{1,2,3}$
 are elements of a complex Clifford algebra
 $\,\{\,\Gamma^M\,,\,\Gamma^{N\,\dagger}\,\}=\d^{MN}$,
 and $\tilde{E}_M\,^m$ is the inverse vielbein.
 For the case at hand, $\tilde{E}_M\,^m$ is given by (\ref{inversedrei}),
 using which we determine
 \brr \psi^1 &=& \Gamma^1
      \nonumber\\[.1in]
      \psi^2 &=& e^{-\phi}\,\bpl\,\Gamma^2
      -\frac{{\rm Re}\,\tau}{{\rm Im}\,\tau}\,
      \Gamma^3\,\bpr
      \nonumber\\[.1in]
      \psi^3 &=&
      \frac{e^{-\phi}}{{\rm Im}\,\tau}\,\Gamma^3 \,.
 \label{qchis}\err
 Substituting (\ref{qchis}) into (\ref{classq}) we obtain
 after a small amount of algebra,
 \brr Q &=& i\,\tilde{\cDslash}
      +e^{-\phi}\,\bpl\,\nu_2\,\Gamma^2
      +\frac{1}{{\rm Im}\,\tau}\,(\,\nu_3
      -{\rm Re}\,\tau\,\nu_2\,)\,\Gamma^3\,\bpr
      \nonumber\\[.1in]
      & & \hspace{.5in}
      +e^\phi\,\bpl\,\
      (\,\mu^2+{\rm Re}\,\tau\,\mu^3\,)\,\Gamma^{2\,\dagger}
      +{\rm Im}\,\tau\,\mu^3\,\Gamma^{3\,\dagger}\,\bpr \,.
 \label{firstq}\err
 By using (\ref{drei}) and (\ref{inversedrei}), we can re-write
 (\ref{firstq}) as
 \brr Q &=& i\,\tilde{\cDslash}
      +\Gamma^M\,\tilde{E}_M\,^m\,\nu_m
      +\mu^m\,E_m\,^M\,\Gamma_M^\dagger
      \nonumber\\[.1in]
      &=&
      i\,\tilde{\cDslash}
      +\Gamma^M\,\nu_M
      +\mu^M\,\Gamma_M^\dagger\,.
 \label{qqqq}\err
 It is gratifying that the quantum supercharge organizes into
 an object with manifest target-space transformation properties.
 The structure of (\ref{qqqq}) also suggests that there is quite
 likely a non-trivial generalization of the $R_i\leftrightarrow 1/R_i$
 duality encountered in the case of the $(S^1)^{D-1}$
 compactification described above.  To investigate this, we will,
 in the next two sections, look at two classes of transformations which
 one can make in the case of the $T^2$ compactifications, each of
 which is codified as a transformation of the torus modulus $\tau$.
 The first class of transformations describes the $SL(2,\Z)$
 modular group describing re-parameterizations of the torus.
 This set of transformations describes an expected symmetry
 group. The second class of transformations includes scale
 transformations.  It is less clear from the basic considerations
 described in this paper that these should comprise a symmetry
 although, as we will show, these do in fact describe a verifiable
 duality relationship.

 Using the Clifford algebra, it is easy to show that the central
 charge operator, defined as $Z=Q^2$, is given by
 \brr Z=\nu_m\,\mu^m \,.
 \err
 Note that this is proportional to the unit operator,
 and is therefore diagonal in any basis.  Note that, based on
 developments to this point, the central charge $Z$
 is not subject to a quantization condition.  This is because
 although the $\nu_i$ are integers, there is no {\it a priori}
 quantization condition on the permitted values of $\mu^i$.
 However, as explained below in section \ref{scale}, a
 $\phi\to -\phi$ duality exists when $\mu^i$ are quantized
 in units of $|\,\tau\,|/{\rm Im}\,\tau=1/\sin\a$,
 where $\a$ is the phase of $\tau$.

 \setcounter{equation}{0}
 \section{Modular Transformations}
 It is interesting to consider the invariance properties of the
 $\R\times T^2$ model by computing what happens to the supercharge
 $Q$ when the parameters describing the torus are modified.
 As is well known, these transformations are described by
 an $SL(2\,,\,\Z)$ group of transformations which acts on the modular
 parameter $\tau$.  In our analysis we will also
 keep careful track of the overall size of our torus.  This is
 facilitated by the real parameter $\phi$, which may be chosen
 independently of the complex modulus $\tau$.
 In this section we consider only transformations
 that preserve the scale of the torus.  Consistency
 requires that the quantum theory is invariant under these.
 One purpose of this section is to demonstrate that this is
 so for size-preserving modular transformations on the $T^2$ factor in
 these compactification schemes.  In the following
 section we will consider certain transformations which do change
 the size of the torus.

 Consider, for example, the re-parametrization
 $R_2\leftrightarrow R_3$, taken along with $\a\to\pi-\a$.
 In terms of $\phi$, ${\rm Re}\,\tau$ and
 ${\rm Im}\,\tau$, this transformation is described by
 \brr T: \hspace{.3in}
      {\rm Re}\,\tau &\to&
      -\frac{1}{|\,\tau\,|^2}\,{\rm Re}\,\tau
      \nonumber\\[.1in]
      {\rm Im}\,\tau &\to&
      \frac{1}{|\,\tau\,|^2}\,{\rm Im}\,\tau
      \nonumber\\[.1in]
      \phi &\to& \phi+\ln |\,\tau\,|
       \,.
 \label{tone}\err
 The transformation of $\phi$ compensates for the scale change
 inherent in the $\tau$ transformations, in such a way that the
 overall size of the torus is maintained.
 We then find that $Q$, given in (\ref{firstq}), is invariant if we also take
 \brr T: \hspace{.3in}
      \ba{c}\nu_2\\\nu_3\ea &\to&
      \ba{cc}&1\\-1&\ea\,
      \ba{c}\nu_2\\\nu_3\ea
      \nonumber\\[.1in]
      \ba{c}\mu^2\\\mu^3\ea &\to&
      \ba{cc}&1\\-1&\ea\,
      \ba{c}\mu^2\\\mu^3\ea
      \nonumber\\[.1in]
      \ba{c}\Gamma^2\\\Gamma^3\ea
      &\to&
      \frac{1}{|\,\tau\,|}\,
      \ba{cc}{\rm Re}\,\tau&{\rm Im}\,\tau\\
      -{\rm Im}\,\tau & {\rm Re}\,\tau\ea\,
      \ba{c}\Gamma^2\\\Gamma^3\ea \,.
 \err
 It is reassuring that we can find a
 transformation on $\mu^i$, $\nu_i$ and $\Gamma^M$ which, in conjunction with
 $R_2\leftrightarrow R_3$, $\a\to\pi-\a$ leaves $H$
 invariant, since this describes nothing more than
 a re-labelling of the coordinates on the $T^2$.

 Next consider the transformation obtained by simply adding
 $2\,\pi$ to the twist angle.  This is given by
 \brr S: \hspace{.3in}
      {\rm Re}\,\tau &\to& {\rm Re}\,\tau+1
      \nonumber\\[.1in]
      {\rm Im}\,\tau &\to& {\rm Im}\,\tau
      \nonumber\\[.1in]
      \phi &\to& \phi
 \label{firsts}\err
 Then $Q$ is invariant if we also take
 \brr S: \hspace{.3in}
      \ba{c}\nu_2\\\nu_3\ea &\to&
      \ba{cc}1&0\\1&1\ea\,
      \ba{c}\nu_2\\\nu_3\ea
      \nonumber\\[.1in]
      \ba{c}\mu^2\\\mu^3\ea &\to&
      \ba{cc}1&-1\\0&1\ea\,
      \ba{c}\mu_2\\\mu_3\ea
      \nonumber\\[.1in]
      \ba{c}\Gamma^2 \\ \Gamma^3\ea &\to&
      \ba{cc}1&0\\0&1\ea\,
      \ba{c}\Gamma^2\\\Gamma^3\ea \,.
 \label{mats}\err
 Again, it is reassuring that we can find a
 transformation on $\mu^i$, $\nu_i$ and $\Gamma^M$ which, in conjunction with
 (\ref{firsts}) leaves $H$ invariant, since this latter transformation is nothing
 more than a re-parametrization of the $T^2$.

 Taken together, the $T$ and $S$ transformations described above
 generate the group $SL(2\,,\,\Z)$.  The generating
 transformations on the complex modulus and on the scale factor
 are
 \brr T:
      && \tau \to -\frac{1}{\tau}
      \hspace{.4in}
      \phi\to\phi+\ln|\,\tau\,|
      \nonumber\\[.1in]
      S(n):
      && \tau \to \tau+n
      \hspace{.3in}
      \phi\to\phi \,,
 \label{stdef}\err
 where $n\in Z$. A generic action is obtained by considering
 $S(\,\frac{b+1}{d}\,)\,T\,S(\,d\,)\,T\,S(\,\frac{1-c}{d}\,)\,T$\,,
 where $b,c,d\in\Z$. Applying these operations right to left on $\tau$ we obtain
 \brr \tau &\to&
      \frac{a\,\tau+b}{c\,\tau+d} \,,
 \err
 where $a\,d-b\,c=1$.
 Now applying using the same sequence of transformations, using the matrices
 appearing in (\ref{mats}), we obtain
 \brr \nu_m &\to& \nu_n\,(\,M^{-1}\,)^n\,_m
      \nonumber\\[.1in]
      \mu^m &\to& (\,M\,)^m\,_n\,\mu^n
 \err
 where
 \brr (\,M\,)^m\,_n=\ba{cc}-a&b\\c&-d\ea\,.
 \err
 Notice that the central charge $Z=\nu_m\,\mu^m$ is
 $SL(2\,,\,\Z)$ invariant.

 We have shown that when these transformations arise from a re-parametrization
 of the $T^2$, they do not alter the theory.
 This provides a useful consistency check, since a mere
 re-parametrization cannot change the physics.

 \setcounter{equation}{0}
 \section{Scale Transformations}
 \label{scale}
 Consider the scale transformation, $R_i\to R_i^{-1}$,
 taken along with $\a\to\pi-\a$.
 In terms of the complex modulus and the scale factor, this
 transformation is described by
 \brr T: \hspace{.3in}
      {\rm Re}\,\tau &\to&
      -\frac{1}{|\,\tau\,|^2}\,{\rm Re}\,\tau
      \nonumber\\[.1in]
      {\rm Im}\,\tau &\to&
      \frac{1}{|\,\tau\,|^2}\,{\rm Im}\,\tau
      \nonumber\\[.1in]
      \phi &\to& -\phi\,.
 \label{cat}\err
 This transformation acts the same way on $\tau$ as the $T$
 transformation given in (\ref{tone}), but acts differently on
 $\phi$.
 This transformation
 is more interesting than the $T$ transformation, however, since it
 exchanges a ``small" torus with a ``large" torus, rather than merely
 re-parameterizing the same torus.
 If we apply the transformations to the supercharge $Q$, given in
 (\ref{firstq}), we find that the supercharge is mapped to its
 Hermitian conjugate $Q\to Q^\dagger$, provided we
 simultaneously transform the parameters $(\,\nu_i\,,\,\mu^i\,)$ and the
 elements of the Clifford algebra according to
 \brr \ba{c}\nu_{2,3}\\\mu^{2,3}\ea &\to&
      \ba{cc}& |\,\tau\,|^{-1}\,{\rm Im}\,\tau\\
      |\,\tau\,|\,(\,{\rm Im}\,\tau\,)^{-1}&\ea\,
      \ba{c}\nu_{2,3}\\\mu^{2,3}\ea
      \nonumber\\[.1in]
      \ba{c}\Gamma^2\\\Gamma^3\ea
      &\to& \frac{1}{|\,\tau\,|}\,
      \ba{cc}{\rm Im}\,\tau&-{\rm Re}\,\tau\\
      {\rm Re}\,\tau&{\rm Im}\,\tau\,\ea\,
      \ba{c}\Gamma^2\\\Gamma^3\ea \,,
 \label{vvv}\err
 along with $\Gamma^1\to\Gamma^{1\,\dagger}$.
 In the case of an orthogonal lattice, where the torus modulus is purely
 imaginary $\tau=i\,|\,\tau\,|$, the transformation (\ref{vvv}) is
 the same as (\ref{dualt}).

 The Hamiltonian is given by
 $H=\ft12\,\{\,Q\,,\,Q^\dagger\,\}$.  Therefore,
 under the transformation given by (\ref{cat}) and (\ref{vvv}), which
 induces $Q\to\Omega^\dagger\,Q^\dagger\,\Omega$,  we have $H\to
 \tilde{H}=\Omega^\dagger\,H\,\Omega$, where $\Omega$
 generates the $\Z_2$ parity automorphism of
 the Clifford algebra described by the transformations
 of $\Gamma^{1,2,3}$ .  Since $H$ and $\tilde{H}$
 are related by a unitary transformation, it follows that $H$ and
 $\tilde{H}$ are iso-spectral, and that the
 transformation represents a duality. Notice also that the supersymmetry central
 charge $Z=\nu_2\,\mu^2+\nu_3\,\mu^3$ is invariant under this
 duality transformation.

 \subsection{Central Charge Quantization}
 Since $\nu_{1,2}\in\Z$ it follows from (\ref{vvv}) that the existence
 of a $\phi(X^1)\to -\phi(X^1)$ duality is contingent upon a quantization
 of $\mu^{1,2}$ as well.  In particular, the duality requires
 \brr \frac{{\rm Im}\,\tau}{|\,\tau\,|}\,\mu^{2,3}\in\Z \,.
 \label{muquant}\err
 Thus, $\mu^{2,3}$ are quantized in units of $1/\sin\a$, where
 $\a$ is the phase of the complex modulus $\tau$.  In the case
 described in section \ref{torrid}, where $\a=\pi/2$, the duality
 is present only if $\mu^{2,3}$ are integers.
 In more general $T^2$ compactifications, the presence of our
 $\phi\to -\phi$ duality
 implies that $Z$ quantization is correlated with the phase of the complex
 modulus $\tau$.

 In summary, provided the parameters $\mu^{2,3}$ are quantized
 according to (\ref{muquant}), it follows that under the
 ``large" $\leftrightarrow$ ``small" torus transformation
 given by
 \brr \tau &\to& -\frac{1}{\tau}
      \nonumber\\[.1in]
      \phi &\to& -\phi \,,
 \err
 the quantized charge operators transform according to
 \brr H &\to& \Omega^\dagger\,H\,\Omega
      \nonumber\\[.1in]
      Q &\to& \Omega^\dagger\,Q^\dagger\,\Omega
      \nonumber\\[.1in]
      Z &\to& \Omega^\dagger\,Z\,\Omega \,.
 \err
 where $\Omega$ generates a $\Z_2$ parity.  We are certain that
 this structure generalizes to much more general compactification
 schemes.  The examples described in this paper provide the
 simplest examples of a more pervasive phenomenon which we hope to
 address more fully in the near future.

 \setcounter{equation}{0}
 \section{Conclusions}
 We have shown explicitly how non-trivial supersymmetry central charges
 are naturally incorporated into quantum mechanical
 sigma models as background vector fields.
 We have explained how these vector fields
 are constrained along with the target space
 metric so as to satisfy a particular
 set of coupled differential equations.  We have explicitly
 quantized models having target-space
 topology $\R\times (S^1)^{D-1}$ and others with
 topology $\R\times T^2$.  In the second class of models we have
 proven the quantum invariance under
 $SL(2\,,\,\Z)$ modular transformations that preserve the size of
 the $T^2$ factor.  In both cases, we have shown the existence of
 a $\Z_2$ duality that equates models with ``large" compact space
 with ostensibly distinct models having ``small" compact spaces.

 The emergence of $T$-duality in the manner
 demonstrated in this paper might be construed as an obvious
 manifestation of known dualities in string theory.  Although we are fairly
 certain that the two classes of phenomena are intimately related, we also
 believe that making a firm connection between $T$-duality in
 string theory and $T$-duality in quantum mechanics is not as
 trivial an exercise as it might superficially seem.  For
 instance, by dimensionally-reducing a two-dimensional
 sigma model, one degenerates the length of the string to
 zero size.  This operation requires that the size of any
 internal cycle which the string wraps also degenerates.
 However, the appearance of $T$-dualities in supersymmetric quantum
 mechanics is insensitive to the size of these cycles.  We think
 it would be interesting to explain the quantization of the
 parameters $\mu^i$ in terms of the topological
 quantization of winding modes in string theory, and plan to address this in
 a future paper.

 As mentioned in the introduction, we hope, among other things, to use
 the constructions in this paper as a basis for further elucidating the
 geometric or topological meaning of shape invariance.
 Typically, shape invariance is explained in terms of an algebraic
 relationship connecting superpotentials in otherwise distinct
 sectors of extended models.  It is possible to use the sigma
 models described in this paper to describe precisely these sorts
 of extended models.  One way to do this is to choose a particular
 matrix representation for the $\Gamma^M$ operators which appear in
 our models.  If one diagonalizes the Hamiltonian, then this
 delineates a multiplicity of sectors, each of which has its own
 superpotential.  This is readily accomplished for the
 $\R\times (\,S^1\,)^{D-1}$ models and $\R\times T^2$ models
 which we have presented.  The form of these
 superpotentials is determined by the choice of the
 function $R_i(X^1)$.  The functions $R_i(X^1)$ can be tuned to
 provide shape invariant quantum mechanics as an effective theory.
 In these constructions, the shape transformation
 is realized geometrically.  But it is not known how the
 requirement of shape invariance is realized as a specific
 geometric or topological restriction on the background.  We think this is an
 interesting problem, and feel that our sigma model
 constructions should provide a powerful context for probing
 a more fundamental explanation for shape invariance.

 Shape invariance is but one application we see for the ideas in this
 paper.  Indeed, the constructions developed in this paper are
 sufficiently basic that we anticipate that they might prove useful
 in a variety of problems in physics.  For instance, in
 \cite{GatesSqm} an operation called automorphic duality is introduced
 which appropriates the notion of Hodge duality into the context
 of quantum mechnanics.  It is found that this operation can be
 performed only on models which exhibit target space isometries.
 We have shown in this paper that this is precisely the
 condition needed to include a supersymmetry central charge vector
 into the background. Since we have also shown that these background
 fields imply interesting target-space dualities, our work
 implies a basic connection between worldline automorphic
 duality and nontrivial target space dualities.  We think that
 this, and related issues, are worthy of further study.

 \appendix

 \setcounter{equation}{0}
 \section{Superfield Conventions}
 \label{conventions}
 \renewcommand{\theequation}{\ref{conventions}.\arabic{equation}}
 In this paper we have used a $d=1$ $N=1$ superspace
 \footnote{We refer to the smallest $d=1$ superspace,
 having one {\it real} anti-commuting coordinate as an $N=1/2$
 superspace.},
 where the $N=1$ implies that
 there is one complex anti-commuting coordinate $\t$.  A general,
 unconstrained superfield is therefore described by
 \brr {\cal S}=A+i\,\t\,\psi+i\,\t^\dagger\,\lambda^\dagger
      +\t^\dagger\,\t\,C
 \err
 where $A$ and $C$ are independent complex commuting component fields,
 and $\psi$ and $\lambda$ are independent complex anti-commuting
 component fields.  Thus, this superfield describes 4+4 off-shell degrees of
 freedom.  The basic supercharge operator $Q_0$ and the superspace
 derivatives $D$ are described by
 \footnote{It should not cause a problem that we have used $D$ for the
 target space dimensionality and also for the superspace
 derivatives; the distinction is naturally clear from the
 contexts in which it is used!}
 \brr Q_0 &=& \frac{\der}{\der\,\t}
      -i\,\t^\dagger\,\der_t
      \hspace{.3in}
      D = \frac{\der}{\der\,\t}
      +i\,\t^\dagger\,\der_t
      \nonumber\\[.1in]
      Q_0^\dagger &=& \frac{\der}{\der\,\t^\dagger}
      -i\,\t\,\der_t
      \hspace{.3in}
      D^\dagger = \frac{\der}{\der\,\t^\dagger}
      +i\,\t\,\der_t \,.
 \label{soup}\err
 We distinguish the basic supercharge operator $Q_0$ from the
 harmonic supercharge operator $Q$ defined in (\ref{scharges})
 by the subscript $0$.  The operator $Q_0$ generates a basic
 supersymmetry transformation on component fields via
 the superspace operation $\d_Q^{(0)}(\e)=\e\,Q_0+\e^\dagger\,Q_0^\dagger$.

 In the main text, we have constructed sigma models involving
 $D$ real superfields $V^1,...,V^D$, where the
 reality of the superfield implies $V^n=V^{n\,\dagger}$.  These
 are defined by
 \brr V^n=X^n+i\,\t\,\psi^n
      +i\,\t^\dagger\,\psi^{n\,\dagger}
      +\t^\dagger\,\t\,B^n \,,
 \err
 where $X^n$ and $B^n$ are real-valued component fields and
 $\psi^n$ are complex anti-commuting component fields.
 Thus, each real multiplet has $2+2$ off-shell degrees of
 freedom, and describes one type of irreducible multiplet.
 An arbitrary differentiable function involving the $V^n$
 is given by
 \footnote{As a simple example of (\ref{funcv}),
 it might be useful to exhibit a quadratic expression,
 \brr V^m\,V^n=
      X^m\,X^n
      +2\,i\,\t\,\,X^{(m}\,\psi^{n)}
      +2\,i\,\t^\dagger\,X^{(m}\,\psi^{n)\,\dagger}
      +\t^\dagger\,\t\,\bpl\,X^{(m}\,B^{n)}
      +\psi^{(m\,\dagger}\,\psi^{n)}\,\bpr\,.
 \nonumber\err}
 \brr F(V) &=& F(X)
      +i\,\t\,F_n(X)\,\psi^n
      +i\,\t^\dagger\,F_n(X)\,\psi^{n\,\dagger}
      \nonumber\\[.1in]
      & &
      +\,\t^\dagger\,\t\,\bpl\,
      F_n(X)\,B^n
      +F_{mn}(X)\,\psi^{m\,\dagger}\,\psi^n\,\bpr
 \label{funcv}\err
 where subscripts on $F_n(V)$ on $F_{mn}(V)$ denote derivatives, e.g.
 $F_n(X)=\der F(X)/\der X^n$. Using the operators
 $D$ and $D^\dagger$ defined in (\ref{soup}), it is simple to compute
 \brr D\,V^n &=&
      i\,\psi^n
      +i\,\t^\dagger\,\bpl\,
      \dot{X}^n+i\,B^n\,\bpr
      -\t^\dagger\,\t\,\dot{\psi^n}
      \nonumber\\[.1in]
      D^\dagger\,V^n &=&
      i\,\psi^{n\,\dagger}
      +i\,\t\,\bpl\,\dot{X}^n-i\,B^n\,\bpr
      +\t^\dagger\,\t\,\dot{\psi}^{n\,\dagger}
 \err
 Note that $(\,D\,V^n\,)^\dagger=-(\,D^\dagger\,V^n\,)$.
 This explains, for instance, why certain superspace expressions
 appearing in the main text, such as (\ref{S0}), are real.

 \subsection{Technique}
 Under a basic supersymmetry transformation, i.e., one which does not
 include the central modifications, a superfield ${\cal S}$ transforms as
 $\d_Q^{(0)}(\e)\,{\cal S}=(\,\e\,Q_0+\e^\dagger\,Q_0^\dagger\,)\,{\cal S}$,
 where $Q_0=\der/\der\t-i\,\t^\dagger\,\der_t$.
 It is straightforward to prove
 \brr \d_Q^{(0)}\,\int\,dt\,d\t\,d\t^\dagger\,(\,-\t\,{\cal S}\,)
      &=&\int\,dt\,d\t\,d\t^\dagger\,(\,\e\,{\cal S}\,)
      \nonumber\\[.1in]
      \d_Q^{(0)}\,\int\,dt\,d\t\,d\t^\dagger\,(\,-\t^\dagger\,{\cal S}\,)
      &=& \int\,dt\,d\t\,d\t^\dagger\,(\,\e^\dagger\,{\cal S}\,)\,.
 \label{nut}\err
 This is easily checked in terms of components.  These expressions
 are valid for any superfield ${\cal S}$, irrespective of whether
 ${\cal S}$ satisfies a reality constraint or any other constraint.  It is
 useful to reverse this argument.  If we are looking for a
 particular superspace expression $\Gamma$ with the property
 $\d_Q^{(0)}\,\Gamma=\int dt\,d\t\,d\t^\dagger\,\e\,{\cal S}$,
 then (\ref{nut}) solves this problem for us.  We see that
 $\Gamma$ is obtained by simply replacing $\epsilon$ with
 $-\theta$.  In other words, $\Gamma=\int
 dt\,d\t\,d\t^\dagger\,(-\t\,{\cal S}\,)$.  Similarly, if we seek
 a superfield expression $\Gamma'$ with the property
 $\d_Q^{(0)}\,\Gamma'=\int dt\,d\t\,d\t^\dagger\,\e^\dagger\,{\cal S'}$.
 then a similar argument tells us $\Gamma'=\int
 dt\,d\t\,d\t^\dagger\,(-\t^\dagger\,{\cal S'}\,)$.
 As a general rule, given a superspace integral which is
 linear in $\e$ or $\e^\dagger$, we obtain an expression whose
 basic supersymmetry variation produces this expression by
 replacing $\e$ with $-\t$ or $\e^\dagger$ with $-\t^\dagger$.

 There is an interesting corollary to this method.
 Suppose we seek superspace expression $\Gamma''$ with the
 property $\d_Q^{(0)}\,\Gamma''=
 \int dt\,d\t\,d\t^\dagger\,\t\,\e\,{\cal S}''$.
 Using our technique, we obtain, by replacing $\e$ with $\t$,
 an expression proportional to $\t^2$, which vanishes identically
 since $\t$ is anti-commuting.
 We conclude that there is no solution to this particular problem;
 that is, the expression $\int dt\,d\t\,d\t^\dagger\,\t\,\e\,{\cal S}''$
 does not describe a basic supersymmetry variation.
 Of course, not every superspace expression linear in $\e$ and
 $\e^\dagger$ is itself a supersymmetry
 variation of another; this is one example. Using a similar
 argument, we see that $\int
 dt\,d\t\,d\t^\dagger\,(\,\e\,\t^\dagger+\e^\dagger\,\t\,)\,{\cal
 S}$ is not a supersymmetry variation, since replacing
 $\e$ with $\t$ and $\e^\dagger$ with $\t^\dagger$ produces
 an expression proportional to the combination
 $(\,\t\,\t^\dagger+\t^\dagger\,\t\,)$, which also vanishes
 identically!   These comments explain some of the technique used in
 section \ref{actionsec}.

 \setcounter{equation}{0}
 \section{Target-space spin structure}
 \label{cders}
 \renewcommand{\theequation}{\ref{cders}.\arabic{equation}}
 In this appendix we assemble some useful relationships pertaining
 to the target space geometry associated with the states in the
 models described in the main text.
 We have considered particles propagating in $D$-dimensional Euclidean
 space.  Accordingly, the local structure group is $SO(D)$.
 Denote coordinate indices using small Latin letters ($m,n,...$),
 structure group indices using capital Latin letters ($M,N,...$),
 and spin indices using small Greek letters ($\a,\b,...$).

 Derivatives covariant with respect to coordinate transformations
 are given by $\nabla_m\,V_n=
 \der_m\,V_n-\Gamma_{mn}\,^l\,V_l$,
 where $\Gamma_{mn}\,^l$ is the affine connection.
 Derivatives covariant with respect to structure group
 transformations are ${\cal D}_m= \der_m
 +\ft12\,\w_m^{MN}\,{\cal O}_{MN}$,
 where $\w_m^{MN}$ is the spin connection
 and ${\cal O}_{MN}$ are $SO(D)$ generators.
 The spin connection is given by
 \brr \w_{m\,MN}=
      E_m\,^P\,\bpl\,\Omega_{MN\,,\,P}
      -2\,\Omega_{P\,[\,M\,,\,N\,]}\,\bpr
 \err
 where $E_m\,^M$ is the vielbein,
 with inverse $\tilde{E}_M\,^m$, and
 $\Omega_{MN\,,\,P}$ is the object of holonomy,
 given by
 \brr \Omega_{MN\,,\,P}
      =-\tilde{E}^m\,_{[M}\,
      \tilde{E}^n\,_{N]}\,
      \der_m\,E_{n\,P}
      \,.
 \err
 In particular, a spin-covariant derivative is given by
 \brr {\cal D}_m\,\psi_\a &=&
      \der_m\,\psi_\a
      +\ft12\,\w_m^{MN}\,(\,\Sigma_{MN}\,)_\a\,^\b\,\psi_\b
      \,,
 \err
 where $\Sigma_{MN}$ generates $SO(D)$ in a spinor
 representation.

 Complex world-line fermions take values in a complex Clifford
 algebra, and transform according to a reducible spinor
 representation of the target space structure group $SO(D)$.
 Accordingly, under an $SO(D)$ transformation
 these transform according to
 \brr \Psi_\a \to
      \bpl\,e^{\ft12\,\t^{MN}\,\Lambda_{MN}}\,\bpr_\a\,^\b\,\Psi_\b
      \,,
 \err
 where $\t^{MN}$ is a real antisymmetric matrix of parameters and
 $\Lambda_{MN}$ are the $SO(D)$ generators in the particular
 reducible spinor representation given by
 \footnote{By way of comparison, real worldline spinors would
 transform according to a smaller $SO(D)$ representation
 described by the generators
 $\Sigma_{MN}=\ft14\,[\,\Gamma_M\,,\,\Gamma_N\,]$,
 where $\Gamma_M=\Gamma_M^\dagger$
 are elements of a real Clifford algebra
 $\{\,\Gamma_M\,,\,\Gamma_N\,\}=2\,\d_{MN}$.
 }
 \brr \Lambda_{MN}=\Gamma_M\,\Gamma^\dagger_N
      -\Gamma_N\,\Gamma^\dagger_M \,,
 \err
 where $\Gamma_M$ are elements of the complex Clifford algebra
 defined by
 $\{\,\Gamma_M\,,\,\Gamma_N^\dagger\,\}=\d_{MN}$
 and by
 $\{\,\Gamma_M\,,\,\Gamma_N\,\}=
 \{\,\Gamma_M^\dagger\,,\,\Gamma_N^\dagger\,\}=0$. It is straightforward
 to prove, using the Clifford algebra, that the generators
 properly represent the $SO(D)$ algebra,
 \brr [\,\Lambda_{MN}\,,\,\Lambda^{OP}\,]=
      -\d_M\,^O\,\Lambda_N\,^P
      +\d_M\,^P\,\Lambda_N\,^O
      -\d_N\,^O\,\Lambda_M\,^P
      +\d_N\,^P\,\Lambda_M\,^O \,.
 \err
 One can easily show that $-\Lambda_{MN}^\dagger$ form another
 reducible representation the same algebra.

 \subsection{A Simple Example}
 Consider a two-dimensional manifold with topology
 $\R\times S^1$.  Parameterize the non-compact dimension
 using $X^1\in\R$ and the compact dimension using an
 angular variable $X^2\in [\,0\,,\,2\,\pi\,]$.
 Choose metric
 $ds^2=(\,dX^1\,)^2+R(X^1)^2\,(\,dX^2\,)^2$, where $R(X^1)$ describes the
 radius of the compact dimension.
 In this case, a possible zweibein is
 \brr E_m\,^M=\ba{cc}1&\\&R(X^1)\ea \,.
 \err
 One then computes the spin connection
 \brr \w_1\,^{12} &=& 0
      \nonumber\\[.1in]
      \w_2\,^{12} &=& -R' \,.
 \err
 Since $X^2$ is an angle, it follows that $\der_2$ is
 quantized according to $i\,\der_2=\nu_2\in\Z$. In this case,
 the spin covariant derivatives are
 \brr {\cal D}_1 &=& \der_1
      \nonumber\\[.1in]
      {\cal D}_2 &=&
      \der_2+\ft12\,\w_2^{MN}\,\Sigma_{MN}
      \nonumber\\[.1in]
      &=& \der_2+\w_2^{12}\,\Sigma_{12}
      \nonumber\\[.1in]
      &=& -i\,\nu_2
      -R'\,\bpl\,
      \Gamma_1\,\Gamma_2^\dagger
      -\Gamma_2\,\Gamma_1^\dagger\,\bpr
      \nonumber\\[.1in]
      &=&
      -i\,\nu_2
      -R'\,\Gamma_1\,\Gamma_2^\dagger
      +R'\,\Gamma_2\,\Gamma_1^\dagger \,,
 \err
 where spin indices have been suppressed.
 Thus,
 \brr \cDslash &=&
      g^{mn}\,{\cal D}_m\,\Gamma_n
      \nonumber\\[.1in]
      &=& g^{mn}\,E_n\,^N\,{\cal D}_m\,\Gamma_N
      \nonumber\\[.1in]
      &=& g^{11}\,E_1\,^1{\cal D}_1\,\Gamma_1
      +g^{22}\,E_2\,^2{\cal D}_2\,\Gamma_2
      \nonumber\\[.1in]
      &=& {\cal D}_1\,\Gamma_1
      +\frac{1}{R^2}\,R\,{\cal D}_2\,\Gamma_2
      \nonumber\\[.1in]
      &=& {\cal D}_1\,\Gamma_1
      +\frac{1}{R}\,{\cal D}_2\,\Gamma_2
      \nonumber\\[.1in]
      &=& \der_1\,\Gamma_1
      +\frac{1}{R}\,\bpl\,
      -i\,\nu_2
      -R'\,\Gamma_1\,\Gamma_2^\dagger
      +R'\,\Gamma_2\,\Gamma_1^\dagger\,\bpr\,\Gamma_2
      \nonumber\\[.1in]
      &=& \der_1\,\Gamma_1
      +\frac{R'}{R}\,\bpl\,
      -\Gamma_1\,\Gamma_2^\dagger\,\Gamma_2
      +\Gamma_2\,\Gamma_1^\dagger\,\Gamma_2\,\bpr
      -i\,\frac{\nu_2}{R}\,\Gamma_2
      \nonumber\\[.1in]
      &=& \der_1\,\Gamma_1
      -\frac{R'}{R}\,
      \Gamma_1\,\Gamma_2^\dagger\,\Gamma_2
      -i\,\frac{\nu_2}{R}\,\Gamma_2
      \nonumber\\[.1in]
      &=& \der_1\,\Gamma_1
      +\ft12\,\frac{R'}{R}\,
      [\,\Gamma_2\,,\,\Gamma_2^\dagger\,]\,\Gamma_1
      -i\,\frac{\nu_2}{R}\,\Gamma_2 \,,
 \err
 where we have used the Clifford algebra, including the relationship
 $\Gamma_2^2=0$.  Thus,
 \brr i\,\cDslash=
      i\,\der_1\,\Gamma_1
      +\ft12\,i\,\frac{R'}{R}\,
      [\,\Gamma_2\,,\,\Gamma_2^\dagger\,]\,\Gamma_1
      +\frac{\nu_2}{R}\,\Gamma_2 \,.
 \err
 This is precisely the relationship which allows us to re-write
 the expression for $Q$ appearing in (\ref{char}) in the manner
 shown in (\ref{limlam}).\\[.2in]

 \noindent
 {\Large{\bf Acknowledgements}}\\[.1in]
 We thank Ted Allen for very useful help in
 organizing our canonical quantization scheme. M.F. is grateful
 to the Slovak Institute for Fundamental Research in
 Podvazie Slovakia where much of this manuscript was prepared.
 D.K. is grateful for an N.S.F. graduate research fellowship.
 We also thank S.J. Gates for pointing out to us various
 connections linking our work with previous discussions
 in the literature.

 \end{document}